\documentclass[prb,twocolumn]{revtex4-1}

\usepackage{graphicx}
\usepackage{float}
\usepackage{amssymb}
\usepackage{amsmath}
\usepackage{dcolumn}
\usepackage{colortbl}
\usepackage{bbold}
\usepackage{array,multirow}
\usepackage{bm}

\renewcommand{\vec}[1]{{\mathbf #1}}

\definecolor{LightGray}{gray}{0.85}
\bibpunct{[}{]}{,}{n}{}{}

\begin{document}

%\preprint{foo}

\title{Ioffe-Regel criterion of Anderson localization in the model of resonant point scatterers}


\author{S.E. Skipetrov}
\email[]{Sergey.Skipetrov@lpmmc.cnrs.fr}
\affiliation{Univ. Grenoble Alpes, CNRS, LPMMC, 38000 Grenoble, France}

\author{I.M. Sokolov}
\email[]{ims@is12093.spb.edu}
\affiliation{Department of Theoretical Physics, Peter the Great St. Petersburg Polytechnic University, 195251 St. Petersburg, Russia}

\date{\today}

\begin{abstract}
We establish a phase diagram of a model in which scalar waves are scattered by resonant point scatterers pinned at random positions in the free three-dimensional (3D) space. A transition to Anderson localization takes place in a narrow frequency band near the resonance frequency provided that the number density of scatterers $\rho$ exceeds a critical value $\rho_c \simeq 0.08 k_0^{3}$, where $k_0$ is the wave number in the free space. The localization condition $\rho > \rho_c$ can be rewritten as $k_0 \ell_0 < 1$, where $\ell_0$ is the on-resonance mean free path in the independent-scattering approximation. At mobility edges, the decay of the average amplitude of a monochromatic plane wave is not purely exponential and the growth of its phase is nonlinear with the propagation distance. This makes it impossible to define the mean free path $\ell$ and the effective wave number $k$ in a usual way. If the latter are defined as an effective decay length of the intensity and an effective growth rate of the phase of the average wave field, the Ioffe-Regel parameter $(k\ell)_c$ at the mobility edges can be calculated and takes values from  {0.3} to  {1.2} depending on $\rho$. Thus, the Ioffe-Regel criterion of localization $k\ell < (k\ell)_c = \mathrm{const} \sim 1$ is valid only qualitatively and cannot be used as a quantitative condition of Anderson localization in 3D.
\end{abstract}

\maketitle

\section{Introduction}
\label{intro}

When the mean free path $\ell$ due to scattering of a wave in a disordered medium becomes of the order of or shorter than the wavelength $\lambda$, a pictorial representation of multiple wave scattering as a sequence of scattering events separated by intervals of free propagation becomes \textit{qualitatively} invalid and new regimes of wave transport may be expected \cite{ioffe60,mott67,mott90,sheng95}. Strictly speaking, the corresponding Ioffe-Regel criterion $k \ell \lesssim 1$, where $k = 2\pi/\lambda$ is the wave number, defines a strong scattering regime in which the nature of wave transport remains to be determined. Because both $k$ and $\ell$ depend on the frequency $\omega$ of the wave, one sometimes defines a Ioffe-Regel frequency $\omega_{\mathrm{IR}}$ solving the equation $k(\omega_{\mathrm{IR}}) \ell(\omega_{\mathrm{IR}}) = \mathrm{const}$, where either $\mathrm{const} = 1$ \cite{sheng94} or  $\mathrm{const} \sim 1$ (e.g.,  $\mathrm{const} = \pi$ in Ref.\ \onlinecite{beltukov13}).
{The Ioffe-Regel frequency $\omega_{\mathrm{IR}}$ separates frequencies $\omega$ for which scattering is weak and elementary excitations in the disordered medium can be viewed, on average, as plane waves that get attenuated as they propagate, from frequencies for which scattering is strong and the elementary excitations cannot be regarded as attenuated plane waves anymore.}

Theory suggests that in a 3D disordered medium, the condition $k \ell = (k\ell)_c \sim 1$ corresponds to a critical point of localization transition {---a mobility edge $\omega_c$} (a metal-insulator transition for electrons in a disordered metal \cite{anderson58}) that separates a region of parameters for which the eigenmodes are spatially extended [$k \ell > (k\ell)_c$] from a region for which the eigenmodes are localized [$k \ell < (k\ell)_c$] \cite{sheng95,anderson58,vollhardt92,evers08}. The Ioffe-Regel criterion $k \ell < (k \ell)_c \sim 1$ is thus often used as a condition to reach Anderson localization of waves in 3D disordered media \cite{kramer93,lagendijk09}. Its importance is difficult to overestimate because it is the only known criterion that makes a link between a transition in macroscopic transport properties and quantities $k$ and $\ell$ that can be calculated microscopically, allowing for engineering of disordered materials with desired localization properties.

Many theoretical works have been devoted to estimations of the critical value $(k\ell)_c$ of the Ioffe-Regel parameter $k\ell$, which inevitably turns out to be close (although not exactly equal) to 1:  $(k\ell)_c = (2 \pi)^{-1/2} \simeq 0.4$ \cite{john84}, 0.84 \cite{economou84}, 0.972 \cite{bart91}, 0.985 \cite{sheng95}.
 {The reason behind such a spread in the values of $(k\ell)_c$ is the absence of exact theory of Anderson localization in 3D. The quantitative estimates of the critical value $(k\ell)_c$ follow from approximate theories: the nonlinear sigma-model \cite{john84}, the so-called potential-well-analogy method \cite{economou84}, and different variants of self-consistent theories in which the diffusion coefficient of a wave is renormalized due to the so-called maximally-crossed diagrams \cite{bart91, sheng95}.} Recent numerical studies of Anderson localization of cold atoms in random optical potentials \cite{yedjour10,piraud14,pasek17} as well as the available ultrasonic \cite{hu08,cobus16} and cold-atom \cite{chabe08,jendr12} experiments are also in agreement with $(k\ell)_c \sim 1$. For light in strongly scattering semiconductor and dielectric powders, values of $k \ell$ down to 2.5 were reported \cite{wiersma97,vanderbeek12,storzer06,sperling13,note1} but an unambiguous signature of Anderson localization has not been convincingly demonstrated \cite{sperling16,skip16a}.

It is worthwhile to note that reaching $k \ell \lesssim 1$ does not ensure Anderson localization. Some models of mechanical vibrations in disordered solids exhibit Anderson transition at frequencies $\omega_c$ that considerably exceed the Ioffe-Regel frequency $\omega_{\mathrm{IR}}$ \cite{sheng94,beltukov13,beltukov17}. Vibrational modes corresponding to frequencies $\omega \in (\omega_{\mathrm{IR}}, \omega_c)$ are extended but cannot be viewed as damped plane waves anymore. Another relevant example is light scattering in an ensemble of identical resonant point scatterers (atoms). Here $k\ell < 1$ can well be reached \cite{fofanov11} but no Anderson localization takes place \cite{skip14,bellando14}.

The present work is devoted to a quantitative verification of the validity of the Ioffe-Regel criterion of localization in a simple model of scalar wave interacting with {\textit{resonant}} point scatterers that are randomly distributed in space. This model exhibits Anderson localization for a band of frequencies $\omega \in (\omega_c^{\mathrm{I}}, \omega_c^{\mathrm{II}})$ near the resonant frequency $\omega_0$ if the number density of scatterers $\rho$ exceeds a critical value $\rho_c$ \cite{skip14, skip16}. We first establish a ``phase diagram'' of the model by determining the mobility edges $\omega_c^{\mathrm{I}}$ and $\omega_c^{\mathrm{II}}$ as functions of scatterer density $\rho$. Next, we calculate the effective wave number $k$ and the mean free path $\ell$ for waves with frequencies exactly at the estimated mobility edges. And finally, we compare their product $(k\ell)_c$ depending on the density $\rho$ with the expected value of 1 and discuss its variations. Our calculated $(k \ell)_c$ varies from  {0.3} to  {1.2} testifying that the Ioffe-Regel criterion $k \ell \leq (k\ell)_c \sim 1$
 {cannot be used as a quantitative criterion of Anderson localization. At the same time, it remains a good qualitative condition of localization, in contrast to the situation encountered in systems with \textit{nonresonant} scattering where the Ioffe-Regel frequency $\omega_{\mathrm{IR}}$ can be very different from the localization transition frequency $\omega_c$ \cite{sheng94,beltukov13,beltukov17}.} As a side but important result, we also demonstrate that the quantities $k$ and $\ell$ start to lose their physical meaning as a localization transition is approached from the extended side because the decay of the absolute value of the ensemble-averaged wave amplitude becomes nonexponential whereas the growth of its phase becomes nonlinear.

\section{Phase diagram for a scalar wave in an ensemble of resonant point scatterers}
\label{diagram}

\begin{figure}[t]
%%\vspace*{-0.5cm}
%%\hspace*{-0.5cm}
\includegraphics[width=0.99\columnwidth]{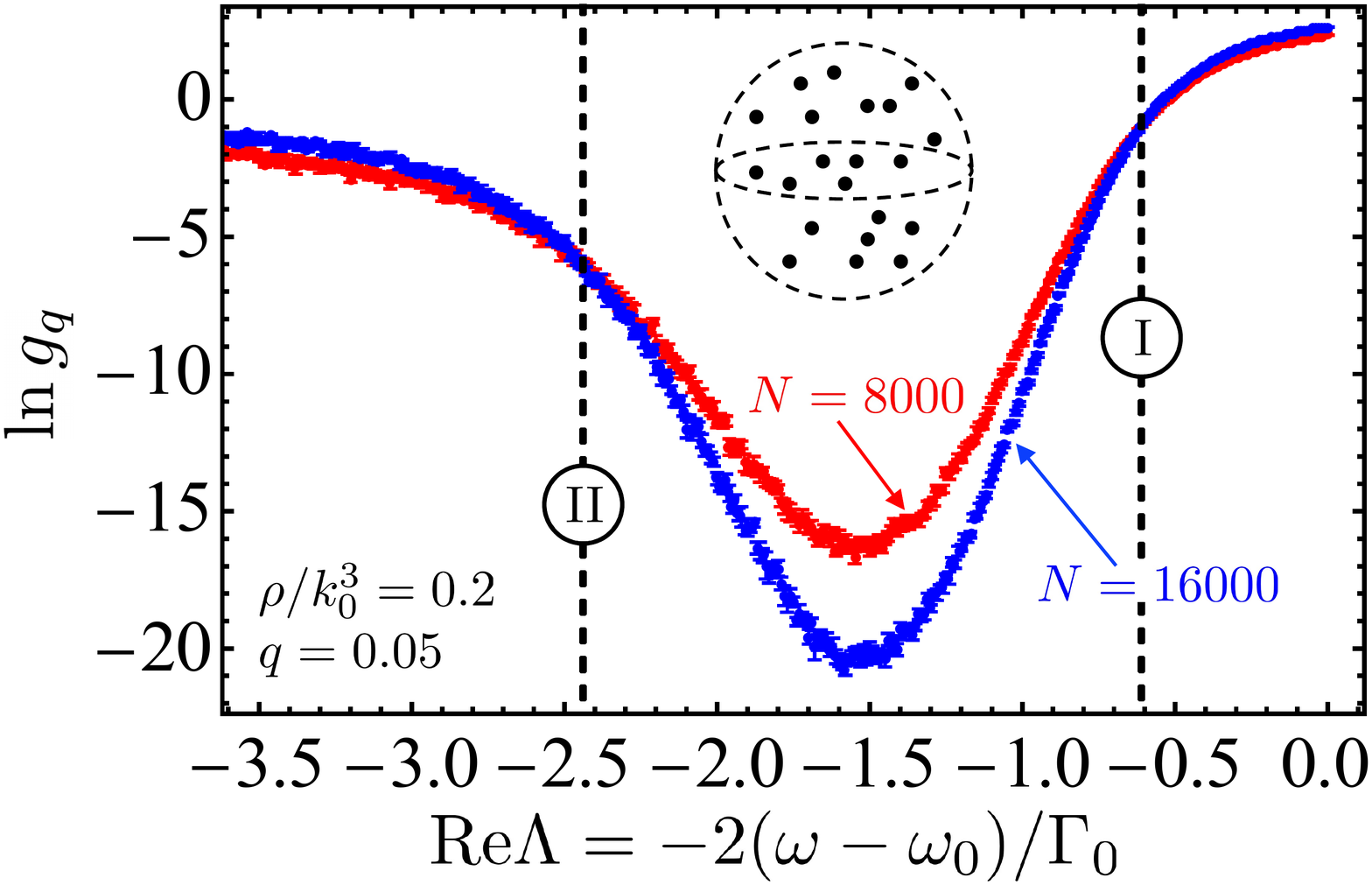}
%%\vspace*{-1.3cm}
\caption{\label{fig_scaling} Illustration of the procedure applied to determine the mobility edges $\omega_c$. Percentiles $\ln g_q$ of the probability distribution of the logarithm of the Thouless conductance $g$ are calculated numerically as functions of the real part $\mathrm{Re} \Lambda$ of the eigenvalues $\Lambda$ of the Green's matrix (\ref{green}), for two different numbers $N$ of point scatterers in a sphere (see the inset), at a fixed density of scatterers $\rho$ (red and blue data points with error bars). The values $\mathrm{Re} \Lambda_c$ of $\mathrm{Re} \Lambda$ at which lines $\ln g_q(\mathrm{Re} \Lambda)$ corresponding to different $N$ cross determine the mobility edges $\omega_c = \omega_0 - (\Gamma_0/2) \mathrm{Re} \Lambda_c$ (dashed vertical lines). Our results are obtained from at least $5.5 \times 10^6$ eigenvalues for each $\rho$ and $N$. Final estimations of $\omega_c$ are averages over results obtained for 10 equispaced values of $q$ in the interval $q = 0.01$--0.1.}
\end{figure}

Consider an ensemble of $N$ identical resonant point scatterers (resonance frequency $\omega_0$, resonance width $\Gamma_0$) randomly distributed inside a spherical volume $V$ of radius $R$
{(see a schematic representation in the inset of Fig.\ \ref{fig_scaling})}.
{The resonant nature of scattering can be due to the quantum internal structure of scatterers, as in the case of a two-level atom with a ground state energy $E_g$ and an excited state energy $E_e$, for which $\omega_0 = (E_e - E_g)/\hbar$ and $\Gamma_0$ is the inverse of the lifetime of the excited state determined by the coupling of the atom to the electromagnetic vacuum (spontaneous emission) \cite{cohen98}. Alternatively, it can be due to classical internal (e.g., sound scattering by small air bubbles in water \cite{ishimaru78}) or geometrical (e.g., Mie scattering of light \cite{bohren04}) resonances. In all cases, realistic scatterers often have multiple scattering resonances and our considerations below apply only to a vicinity of a single resonance that is sufficiently well separated from the others. The point-scatterer assumption implies that the scatterer size is much smaller than the wavelength. It is an excellent approximation for light scattering by atoms (because the latter are indeed much smaller than the optical wavelength) or sound scattering by sufficiently small air bubbles in water, but would be a very crude simplification in the case of Mie scattering for which our model cannot be rigourously justified.}

Multiple scattering of a quasiresonant scalar wave in such a scattering medium can be studied by analyzing the properties of the so-called  ``Green's matrix'' ${\hat G}(\omega_0)$ with elements
\begin{eqnarray}
G_{mn}(\omega_0) = i \delta_{mn} + \left( 1-\delta_{mn} \right) \frac{\exp(i k_0 |\vec{r}_m - \vec{r}_n|)}{k_0 |\vec{r}_m - \vec{r}_n|},
\label{green}
\end{eqnarray}
where  {$\{ \vec{r}_m \}$ are scatterer positions ($m = 1,\ldots,N$)}, $k_0 = \omega_0/c$ and $c$ is the speed of the wave in the absence of scatterers. This approach was first proposed by Foldy \cite{foldy45}, then discussed in detail by Lax \cite{lax51}, and later used in the context of Anderson localization by several authors \cite{skip14,bellando14,skip16,rusek00,pinheiro04}. A transition to Anderson localization upon increasing the number density of scatterers  {$\rho = N/V$} in this model was demonstrated in Ref.\ \onlinecite{skip14} and the critical parameters of this transition were calculated in Ref.\ \onlinecite{skip16}. Here we apply the approach developed in the latter work to determine the critical frequencies $\omega_c$ of the Anderson transitions as functions of $\rho$.  {For consistency of presentation, we provide below a short description of our approach and refer the reader to Ref.\ \onlinecite{skip16} for more details.}

\begin{figure}[t]
%%\vspace*{-0.5cm}
%%\hspace*{-0.5cm}
\includegraphics[width=0.99\columnwidth]{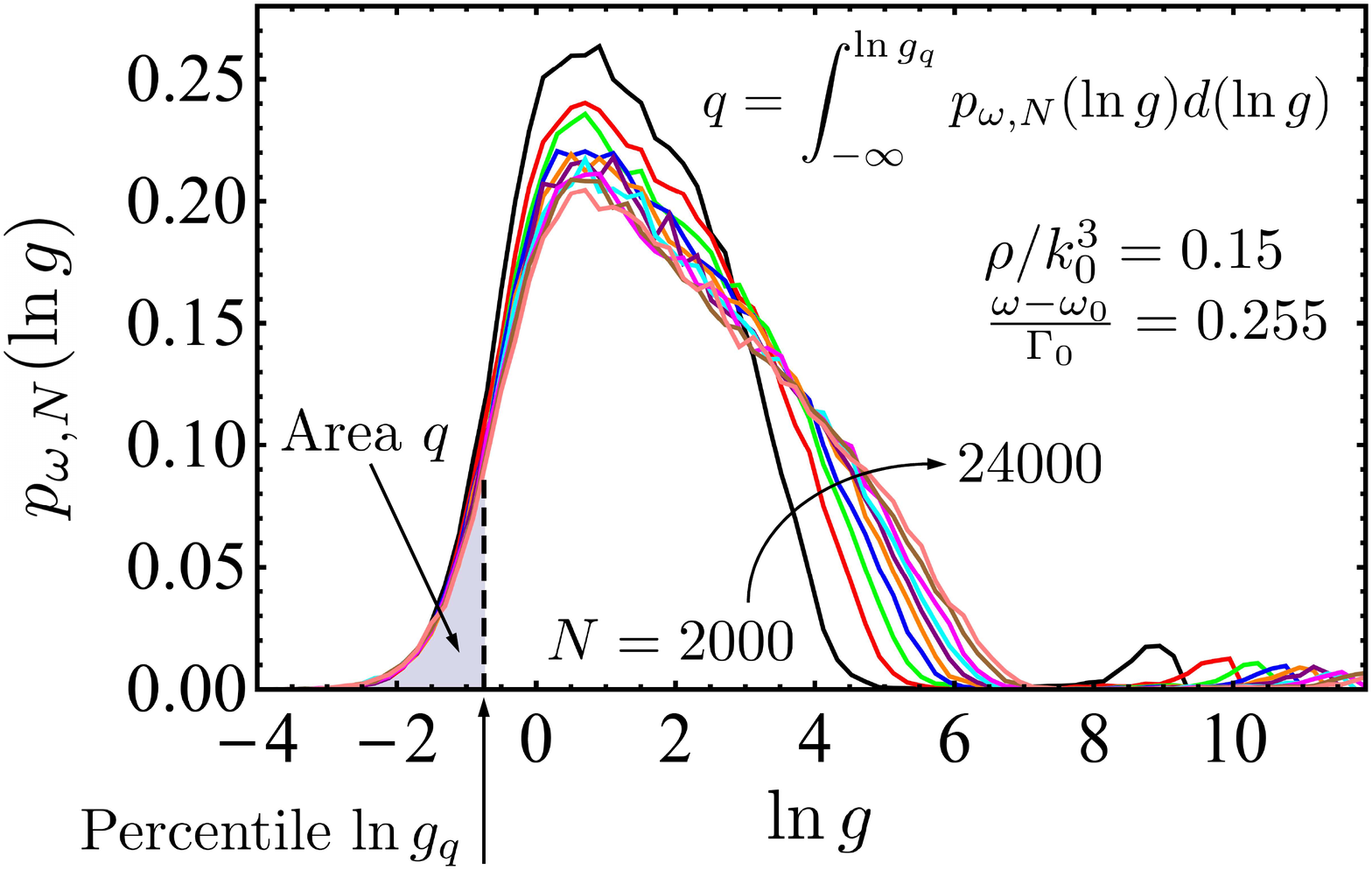}
%%\vspace*{-1.3cm}
\caption{\label{fig_distribution}  {Probability densities $p_{\omega,N}$ of the logarithm of Thouless conductance $g$ at a fixed density of scatterers $\rho/k_0^3 = 0.15$, for a fixed frequency $\omega = \omega_0  + 0.255 \Gamma_0$ close to the critical frequency $\omega_c^{\mathrm{I}}$ of the localization transition, and for different numbers of scatterers $N$. A $(q \times 100)$-th percentile is defined as a limit of integration $\ln g_q$ (shown by a vertical arrow) up to which $p_{\omega,N}(\ln g)$ should be integrated to obtain $q$ (equal to the grey area in the figure) as a result. The illustration in the figure is for the fifth percentile ($q = 0.05$).}}
\end{figure}

 {For a given scatterer number density $\rho$ and a given configuration $\{ \vec{r}_m \}$ of $N$ scatterers, we diagonalize the matrix  ${\hat G}(\omega_0)$ and} define the Thouless conductance $g(\omega_n, N)$ as a ratio of the imaginary part of a  {complex} eigenvalue $\Lambda_n$ to the average distance between projections of $\Lambda_n$'s on the real axis:
\begin{eqnarray}
g(\omega_n, N) = \frac{\mathrm{Im} \Lambda_n}{\langle \mathrm{Re} \Lambda_n - \mathrm{Re} \Lambda_{n-1} \rangle},
\label{thouless}
\end{eqnarray}
where the eigenvalues $\Lambda_n$ are ordered in the order of ascending real parts ($\mathrm{Re} \Lambda_n > \mathrm{Re} \Lambda_{n-1}$) and $\omega_n = \omega_0 - (\Gamma_0/2) \mathrm{Re} \Lambda_n$.
 {The definition (\ref{thouless}) of $g$ follows the spirit of fundamental works \cite{edwards72,thouless74,abrahams79} on Anderson localization in which $g$ was introduced as a measure of sensitivity of a disordered quantum (or, more generally, wave) system to a modification of boundary conditions. The imaginary part of an eigenvalue $\Lambda_n$ is equal to the decay rate (in units of $\Gamma_0$) of the corresponding eigenstate due to the openness of the medium, so that Eq.\ (\ref{thouless}) corresponds to the ``Thouless number'' used in the scaling theory of localization \cite{abrahams79}.} For scatterers at random positions, $g$ becomes a random number with a probability density $p_{\omega,N}(g)$ that we estimate numerically.
 {The single-parameter scaling hypothesis implies that at a critical point of localization transition $\omega = \omega_c$, $p_{\omega,N}(g)$ takes a universal, $N$-independent form \cite{shapiro86,slevin01,slevin03}, provided that $N$ is large enough. In particular, such an universality can be used to identify the mobility edges $\omega_c$. However, we do not find it at any frequency $\omega$. Our analysis shows that only the small-$g$ part of $p_{\omega,N}(g)$ becomes independent of $N$ at two frequencies $\omega_c^{\mathrm{I}}$ and $\omega_c^{\mathrm{II}}$ that we identify as mobility edges (see Fig.\ \ref{fig_distribution}). Such a partial universality may be due either to a breakdown of single-parameter scaling in our system or, which is more likely, to an insufficient number of scatterers $N$ in our calculations, preventing us from reaching the single-parameter scaling regime for the entire probability density function of $g$. Indeed, a careful examination of Fig.\ \ref{fig_distribution} shows a tendency of convergence of $p_{\omega,N}(g)$ towards some limiting distribution with increasing $N$, but the convergence is clearly not achieved yet for $g \gtrsim 1$.}

 {The $N$-independence of the small-$g$ part of $p_{\omega,N}(\ln g)$ (typically, for $\ln g < 0$) at the mobility edges together with a single-parameter scaling hypothesis for this part of the distribution, allows us to apply to it the procedure of finite-size scaling in order to determine the mobility edges accurately. The finite-size scaling analysis of a probability density is more conveniently performed in terms of percentiles $\ln g_q$ of the distribution \cite{slevin03}. The latter are defined by
\begin{eqnarray}
q = \int\limits_{-\infty}^{\ln g_q(\omega,N)} p_{\omega,N}(\ln g) d(\ln g).
\label{perc}
\end{eqnarray}
Figure \ref{fig_distribution} illustrates this definition. The $(q \times 100)$-th percentile is simply the limit of integration up to which the distribution should be integrated to obtain $q$ as a result of integration. The 50-th percentile is the median of the distribution. Although we will only consider the distribution $p_{\omega,N}(\ln g)$ of the logarithm of $g$, it is worthwhile to note that
\begin{eqnarray}
q &=& \int\limits_{-\infty}^{\ln g_q(\omega,N)} p_{\omega,N}(\ln g) d(\ln g)
\nonumber \\
&=& \int\limits_{0}^{g_q(\omega,N)} {\tilde p}_{\omega,N}(g) dg,
\label{perc2}
\end{eqnarray}
where ${\tilde p}_{\omega,N}(g)$ is the probability density of $g$.}

\begin{figure}[t]
%%\vspace*{-0.5cm}
%%\hspace*{-0.5cm}
\includegraphics[width=0.99\columnwidth]{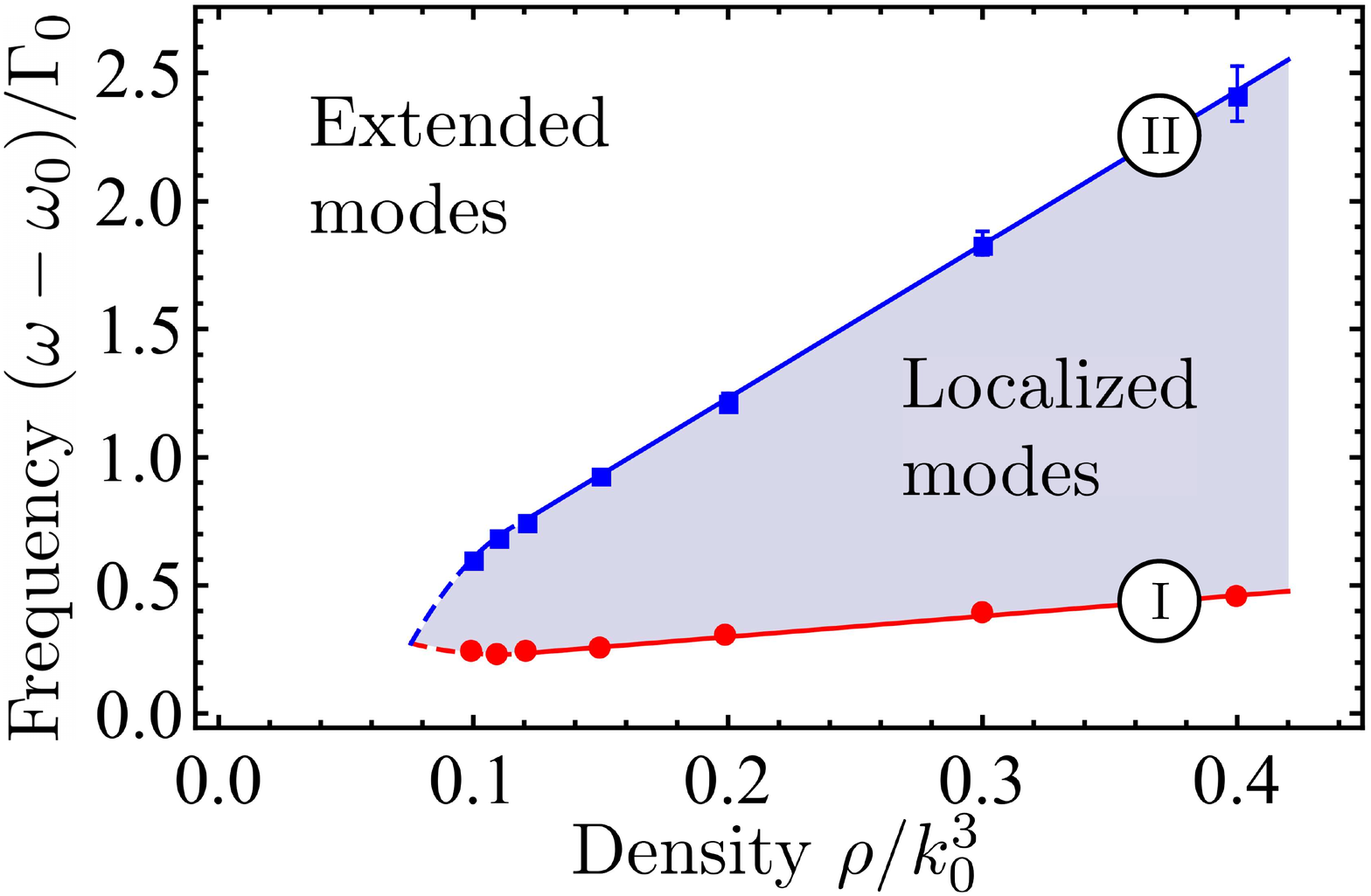}
%%\vspace*{-1.3cm}
\caption{\label{fig_me} Phase diagram of the model of resonant point scatterers: grey and white areas correspond to the regions of localized and extended modes, respectively. Data of Table \ref{tabme} are shown by full circles and squares for the mobility edges I and II, respectively. Straight solid lines are linear fits showing the estimated mobility edges as functions of scatterer density for $\rho/k_0^3 \geq 0.121$. Dashed lines are quadratic fits for $\rho/k_0^3 \leq 0.121$. Their crossing point identifies $\rho_c/k_0^3 \simeq 0.08$ as an estimation of the lowest density at which localized modes may appear.}
\end{figure}

The $N$-independent form of $p_{\omega,N}(\ln g)$ at small $g$ translates into $N$-independent values of $\ln g_q(\omega,N)$ for small $q$. Therefore, the mobility edges $\omega_c$ can be determined by computing $\ln g_q(\omega,N)$ as functions of $\omega$ for several values of $N$ and finding the frequencies $\omega = \omega_c$ for which the curves obtained for different $N$ all cross. The fact that such crossing points indeed exist has been demonstrated in Ref.\ \onlinecite{skip16}, which can be consulted for more details. It is therefore sufficient to use  {any} two different values of $N$ to determine $\omega_c$. This is illustrated in Fig.\ \ref{fig_scaling} for $\rho/k_0^3 = 0.2$ and  $q = 0.05$. The final estimations of $\omega_c$ are obtained by repeating calculations for 10 equispaced values of $q$ in the interval $q = 0.01$--0.1 and averaging over $q$.
 {This interval of $q$ was chosen to limit statistical errors of the calculated mobility edges $\omega_c$, on the one hand, and to restrict the integration in Eq.\ (\ref{perc}) to small $\ln g < 0$, where $p_{\omega,N}(\ln g)$ is assumed to obey the single-parameter scaling, on the other hand. When averaging over a given number of statistically independent realizations of disorder, the errors are larger for smaller $q$ and become too important for $q < 0.01$ and the limited number of realizations that we used. On the other hand, $q > 0.1$ cannot be used because it would rely on $p_{\omega,N}(\ln g)$ for large $\ln g$ for which the single-parameter scaling assumption breaks down, as we discussed above. The critical frequencies $\omega_c$ obtained from our analysis} for 7 different densities $\rho/k_0^3 = 0.1$, 0.11, 0.121, 0.15, 0.2, 0.3 and 0.4 are given in Table \ref{tabme}.

\begin{table}[t]
\caption{\label{tabme} Mobility edges $(\omega_c - \omega_0)/\Gamma_0$.}
\begin{ruledtabular}
\begin{tabular}{lcc}
Density $\rho/k_0^3$ & Mobility edge I & Mobility edge II\\
\colrule
~~~0.1   & $0.234 \pm 0.004$  & $0.608 \pm 0.005$\\
~~~0.11  & $0.231 \pm 0.004$  & $0.693 \pm 0.006$\\
~~~0.121 & $0.236 \pm 0.004$  & $0.753 \pm 0.014$\\
~~~0.15  & $0.256 \pm 0.003$  & $0.935 \pm 0.011$\\
~~~0.2   & $0.305 \pm 0.003$  & $1.219 \pm 0.030$\\
~~~0.3   & $0.385 \pm 0.005$  & $1.836 \pm 0.047$\\
~~~0.4   & $0.451 \pm 0.006$  & $2.419 \pm 0.107$\\
\end{tabular}
\end{ruledtabular}
\end{table}

\begin{figure}[t]
%%\vspace*{-0.5cm}
%%\hspace*{-0.5cm}
\includegraphics[width=0.99\columnwidth]{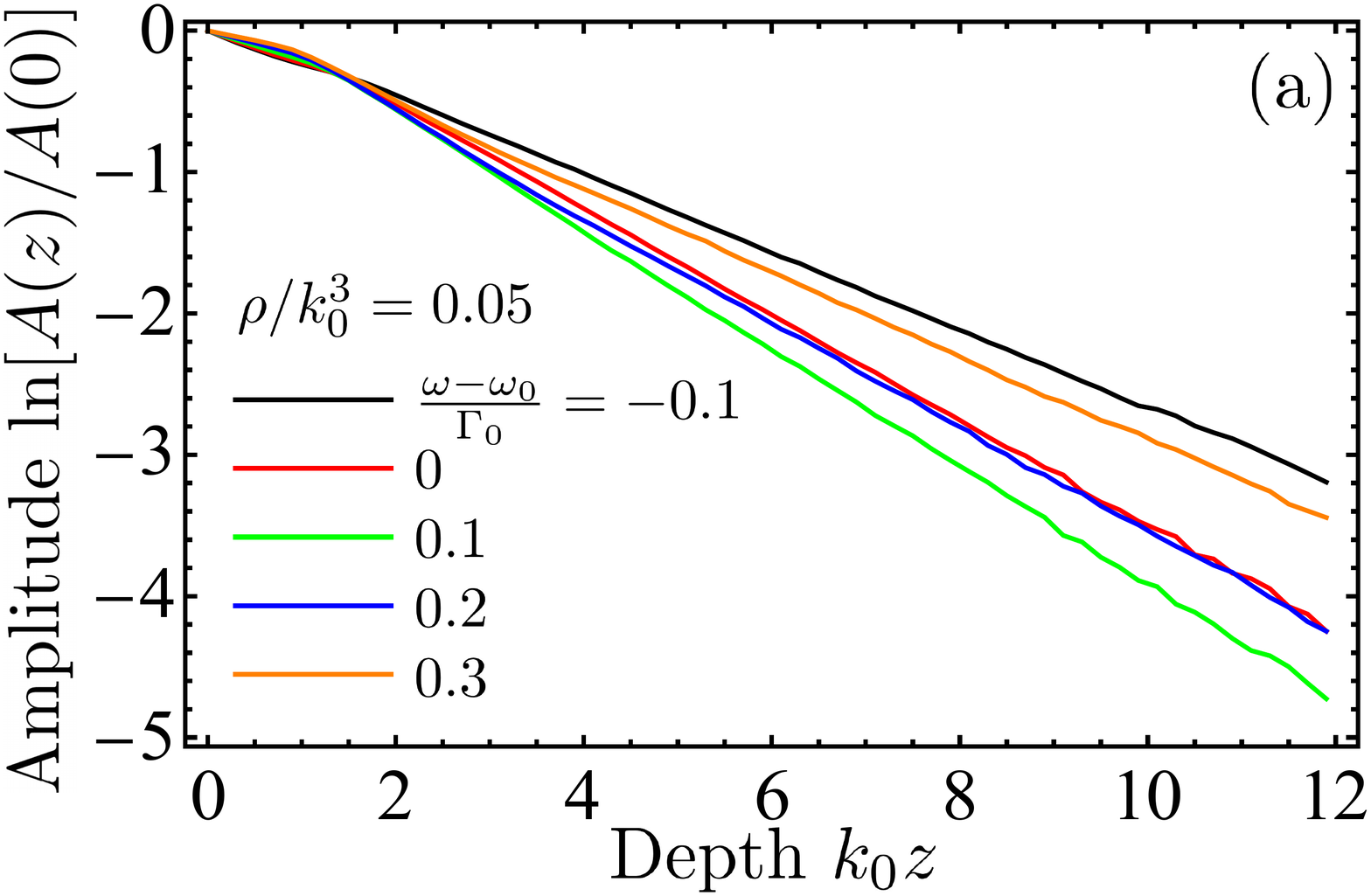}
%% \;\;\;\;
\includegraphics[width=0.99\columnwidth]{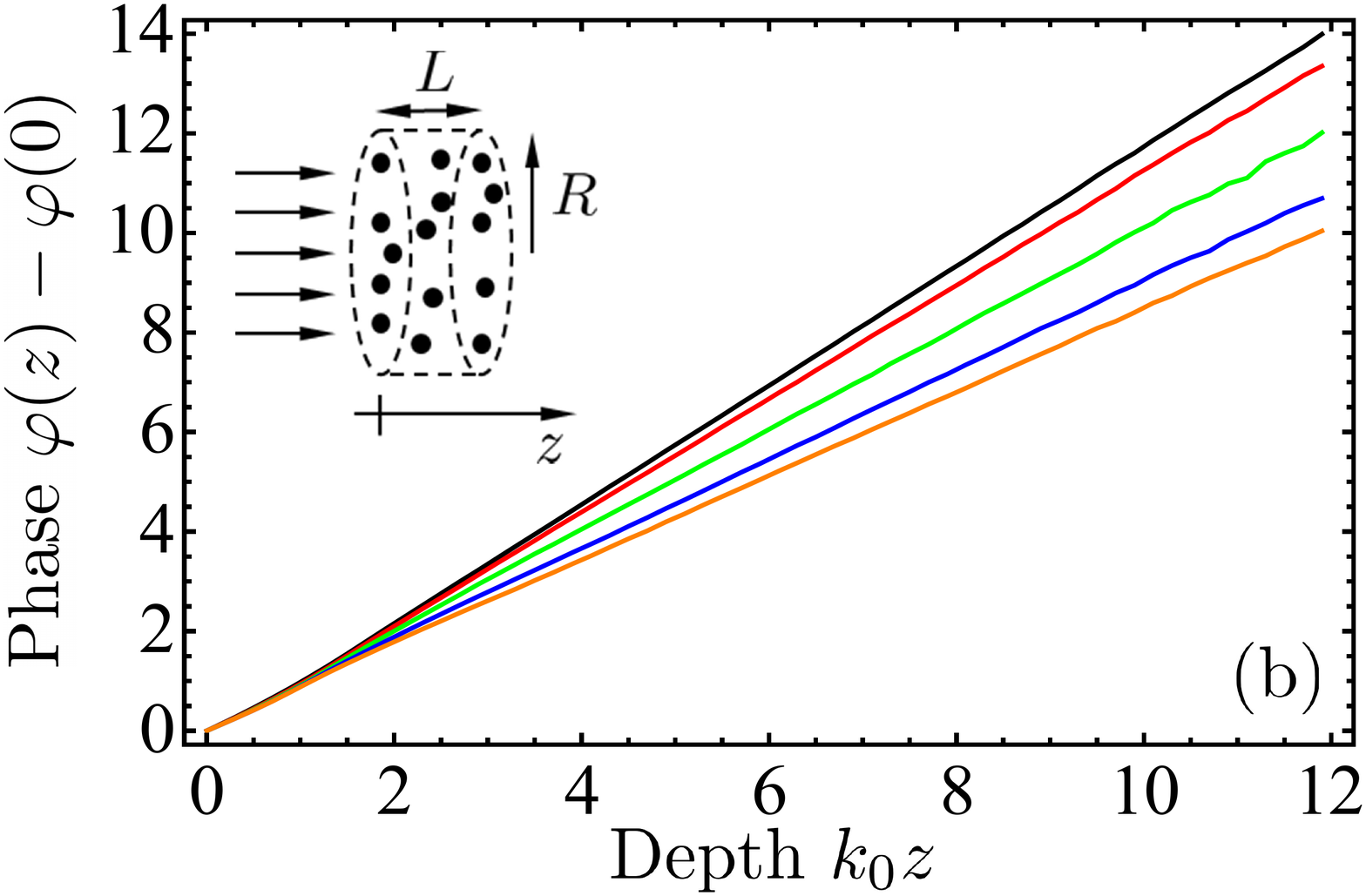}
%%\vspace*{-1.3cm}
\caption{\label{fig_lowrho}  {The exponential decay of the amplitude (a) and the linear growth of the phase (b) of the average wave field as a function of depth $z$ into the medium at a low scatterer number density $\rho/k_0^3 = 0.05$, for several frequencies $\omega$. Averaging is performed over $10^6$ scatterer configurations for each curve. The inset of panel (b) shows the geometry of a cylindrical layer of radius $k_0 R = 20$ and thickness $k_0 L = 12$, illuminated by a monochromatic plane wave, used for the calculations.}}
\end{figure}

Figure \ref{fig_me} represents the data of Table \ref{tabme} graphically (symbols with error bars). Such a representation helps to recognize that both $\omega_c^{\mathrm{I}}$ and $\omega_c^{\mathrm{II}}$ exhibit a roughly linear dependence on $\rho/k_0^3$ for $\rho/k_0^3 \gtrsim 0.12$ (straight solid lines in Fig.\ \ref{fig_me}). The density dependencies of $\omega_c^{\mathrm{I,II}}$ bend at lower densities and we fit the first three data points of each dependence by quadratic polynomials (dashed lines in Fig.\ \ref{fig_me}). The crossing point of the latter identifies $\rho_c/k_0^3 \simeq 0.08$ as a minimal density at which localized modes may appear. A more accurate determination of $\rho_c$ is complicated by the fact that it becomes difficult to determine the values of $\omega_c$ accurately when $\rho$ approaches $\rho_c$. Because $\ell_0 = k_0^2/4\pi\rho$ is the on-resonance mean free path in the independent-scattering approximation, $\rho/k_0^3 = 0.08$ corresponds to $k_0 \ell_0 = 1$ with a high degree of accuracy.

\section{Analysis of the coherent wave field}
\label{coherent}

\begin{figure*}[t]
%%\vspace*{-0.5cm}
%%\hspace*{-0.5cm}
\includegraphics[width=0.96\textwidth]{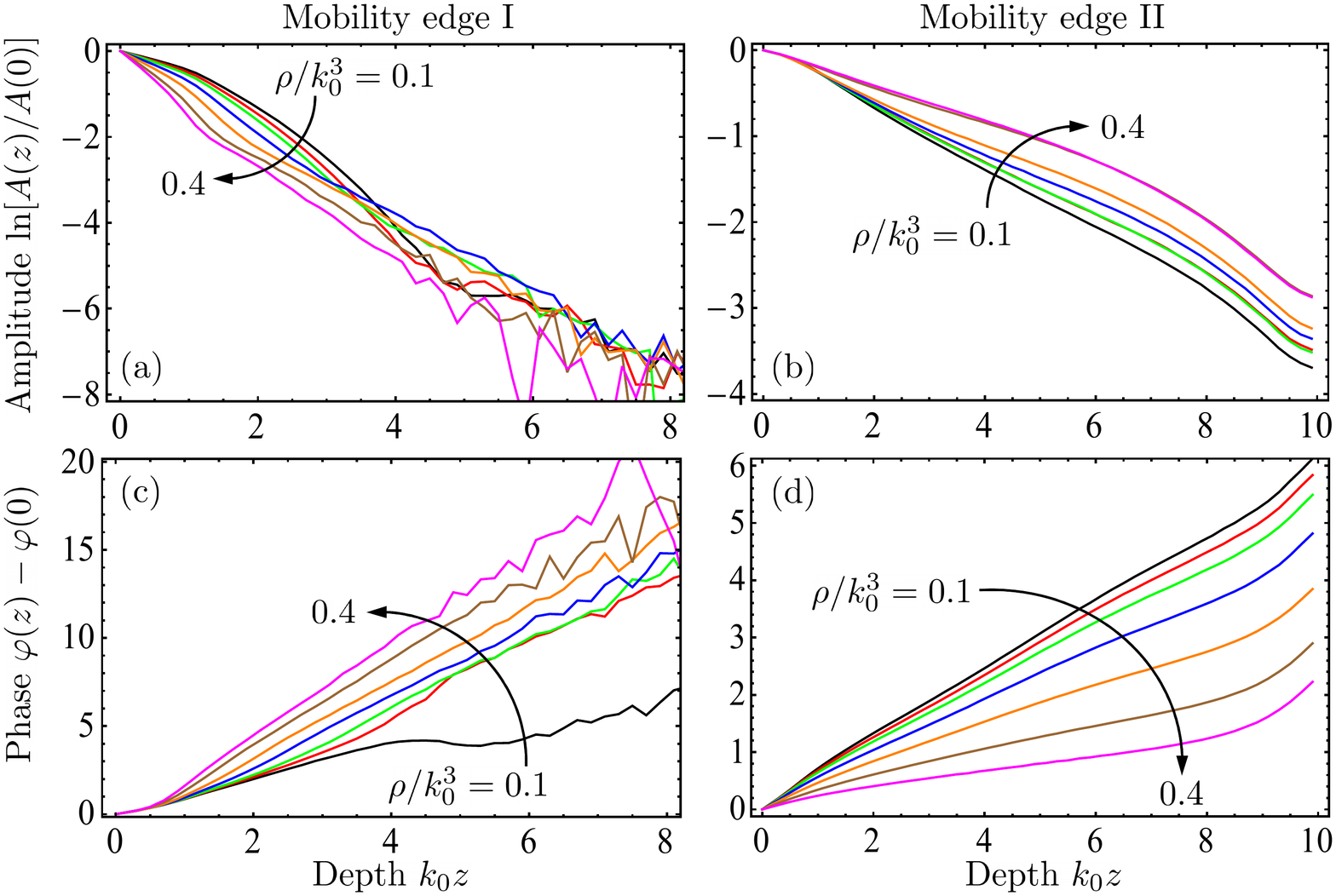}
\vspace*{-0.5cm}
\caption{\label{fig_ampl_phase} Decay of the amplitude [first row, panels (a) and (b)] and growth of the phase [second row, panels (c) and (d)] of the average wave field as a function of depth $z$ into the medium at the mobility edges I (first column) and II (second column) for 7 scatterer densities from $\rho/k_0^3 = 0.1$ to 0.4 listed in Table \ref{tabme}. Averaging is performed over $10^5$--$10^6$ scatterer configurations for each curve.  {These results are obtained in the same geometry as those of Fig.\ \ref{fig_lowrho} [see the inset of Fig.\ \ref{fig_lowrho}(b)] but using $k_0 L = 10$.}}
\end{figure*}

Having found the positions of mobility edges as functions of the scatterer number density, we now want to determine the effective wave number $k$ and the mean free path $\ell$ at the critical points. To this end, we analyze the coherent component of a plane wave incident on an ensemble of resonant point scatterers as it propagates into the sample.
%%We use a scalar version of the approach developed in Ref.\ \onlinecite{sokolov11}.
We consider point scatterers that are randomly distributed inside a cylindrical layer of radius $k_0 R = 20$ and thickness $k_0 L =  {12}$, see the inset of Fig.\ \ref{fig_lowrho}(b). A monochromatic plane wave $\psi_0(\vec{r}) = \exp(i k z)$ with $k = \omega/c$ is incident on the sample from the left, at $z = 0$. A vector $\bm{\psi} = (\psi_1, \ldots, \psi_N)^{T}$ of wave amplitudes $\psi_m = \psi(\vec{r}_m)$ at the points $\{ \vec{r}_m \}$ where the scatterers are located obeys \cite{foldy45,lax51}
\begin{eqnarray}
\bm{\psi} = \bm{\psi}_0 + \alpha(\omega) \left[ {\hat G}(\omega) - i \mathbb{1} \right] \bm{\psi},
\label{foldylax}
\end{eqnarray}
where $\bm{\psi}_0 = [\psi_0(\vec{r}_1), \ldots, \psi_0(\vec{r}_N)]^{T}$ and $\alpha(\omega) = -(\Gamma_0/2)/(\omega - \omega_0 + i \Gamma_0/2)$ is the (dimensionless) scatterer polarizability. The solution of Eq.\ (\ref{foldylax}) reads
\begin{eqnarray}
\bm{\psi} = \left( \mathbb{1} - \alpha(\omega) \left[ {\hat G}(\omega) - i \mathbb{1} \right] \right)^{-1} \bm{\psi}_0.
\label{sol}
\end{eqnarray}
We use this expression to calculate $\bm{\psi}$ for many different random configurations $\{ \vec{r}_m \}$ of scatterers. The results are then averaged over all configurations and over the central part of the cylinder $\sqrt{x^2 + y^2} < 12$ to obtain an average wave field $\langle \psi(z) \rangle$ as a function of penetration depth $z$. $\langle \psi(z) \rangle$ is a complex function and can be represented as $\langle \psi(z) \rangle = A(z) \exp[i \varphi(z)]$, with real amplitude $A(z)$ and phase $\varphi(z)$. From the multiple scattering theory \cite{sheng95,akkermans07} we expect
\begin{eqnarray}
A(z) &=& A(0) \exp(-z/2\ell),
\label{az}
\\
\varphi(z) &=& \varphi(0) + kz,
\label{phiz}
\end{eqnarray}
which define the scattering mean free path $\ell$ and the effective wave number $k$.

The exponential decay of $A(z)$ and the linear growth of $\varphi(z)$ according to Eqs.\ (\ref{az}) and (\ref{phiz}) are in agreement with our calculations  {for all frequencies $\omega$ at sufficiently low densities $\rho/k_0^3 \lesssim 0.05$, see Fig.\ \ref{fig_lowrho}. At higher densities,} Eqs.\ (\ref{az}) and (\ref{phiz})  {start to} break down  {in a narrow frequency band above the single-scatterer resonance frequency $\omega_0$. Deviations from Eqs.\ (\ref{az}) and (\ref{phiz}) become violent} at densities needed to reach Anderson localization. Figure \ref{fig_ampl_phase} shows the dependencies $A(z)$ and $\varphi(z)$ that we find at the mobility edges determined in Sec.\ \ref{diagram}. The amplitude $A(z)$ exhibits a nonexponential decay with $z$ whereas the growth of the phase $\varphi(z)$ is not linear. This is particularly pronounced at the low-frequency mobility edge I (panels of the first row in Fig.\ \ref{fig_ampl_phase}), but is also the case at the high-frequency mobility edge II (panels of the second row in Fig.\ \ref{fig_ampl_phase}).
 {We also note that the decay of $A(z)$ with the depth $z$ is much faster at the first mobility edge than at the second one [compare the scale of the vertical axes in Figs.\ \ref{fig_ampl_phase}(a) and (b)]. One of the consequences of the fast decay of $A(z)$ at the first mobility edge is a bad convergence of our calculations for $k_0 z > 4$, leading to a noisy behavior of the amplitude and phase in Figs.\ \ref{fig_ampl_phase}(a) and (c).}

The nonexponential decay of $A(z)$ and the nonlinear growth of $\varphi(z)$ at large scatterer densities can be traced back to the spatial dispersion of the self-energy $\Sigma$, which  becomes momentum-dependent. Indeed, at low densities the self-energy is proportional to the scattering matrix $t(\omega) = -(4\pi/k_0)\alpha(\omega)$ of an isolated scatterer: $\Sigma(\omega) = \rho t(\omega)$, and it is independent of the momentum $\vec{q}$. The Fourier transform of the average Green's function $G(\vec{q}, \omega) = [\omega^2/c^2 - q^2 - \Sigma(\omega)]^{-1}$ yields $G(\vec{r}, \omega) = -\exp(ikr-r/2\ell)/4\pi r$, where $k = \omega/c - \mathrm{Re} \Sigma(\omega)/2 (\omega/c)$ and $\ell = -(\omega/c)/\mathrm{Im} \Sigma(\omega)$. In its turn, the integration of $G(\vec{r}, \omega)$ over the input surface of a disordered layer to model an incident plane wave yields Eqs.\ (\ref{az}) and (\ref{phiz}). It is known, however, that second-order in $\rho$ corrections to $\Sigma$ are momentum-dependent and $\Sigma = \Sigma(\vec{q}, \omega)$ \cite{bart94,cherroret16}. Including these corrections into the analysis produces deviations from Eqs.\ (\ref{az}) and (\ref{phiz}) that are qualitatively similar to those in our Fig.\ \ref{fig_ampl_phase}. The deviations that we observe are, however, much stronger than those that can be understood using terms up to second order in $\rho$. They cannot be fitted to the theory of Ref.\ \onlinecite{bart94}. Higher-order terms are required to provide a quantitative analytical description of our numerical results {, but properly taking them into account is a formidable task that neither us nor others managed to perform up to now.
One of the peculiarities that cannot be even qualitatively understood using only the low-order terms of the perturbation theory in $\rho/k_0^3 \ll 1$ is the abrupt change in behavior of the phase in Fig.\ \ref{fig_ampl_phase}(c) when increasing the density from  $\rho/k_0^3 = 0.1$ to  $\rho/k_0^3 = 0.11$ (the two lower curves). Further work is needed to clarify the physical reasons behind such a behavior.}

Strictly speaking, the complicated dependencies $A(z)$ and $\varphi(z)$ in Fig.\ \ref{fig_ampl_phase} make it impossible to define the mean free path $\ell$ and the effective wave number $k$. We note, however, that even though the behavior of $\ln A(z)$ and $\varphi(z)$ is nonlinear, they are still monotonically decreasing and increasing functions, respectively. Moreover, even though the decay of $A(z)$ is not purely exponential, it does not slow down considerably and does not become power-law. We thus can define $\ell$ an $k$ as some effective decay length of $A(z)^2$ and growth rate of $\varphi(z)$, respectively. We propose to do it in two ways. First, we introduce integral definitions
\begin{eqnarray}
\ell &=& \int\limits_0^{\infty} \left[ A(z)/A(0) \right]^2 dz,
\label{integralell}
\\
k &=& \frac{2}{z_{\mathrm{max}}^2} \int\limits_0^{z_{\mathrm{max}}} \left[ \varphi(z) - \varphi(0) \right]  dz,
\label{integralk}
\end{eqnarray}
where $z_{\mathrm{max}}$ is a depth up to which we consider our results reliable. It is easy to verify that Eqs.\ (\ref{az}) and (\ref{phiz}) obey Eqs.\ (\ref{integralell}) and (\ref{integralk}), so that the latter will give correct results in the limit of weak disorder. Their advantage is that they will yield physically meaningful results at strong disorder as well.

\begin{figure}[t]
%%\vspace*{-0.5cm}
%%\hspace*{-0.5cm}
\includegraphics[width=0.99\columnwidth]{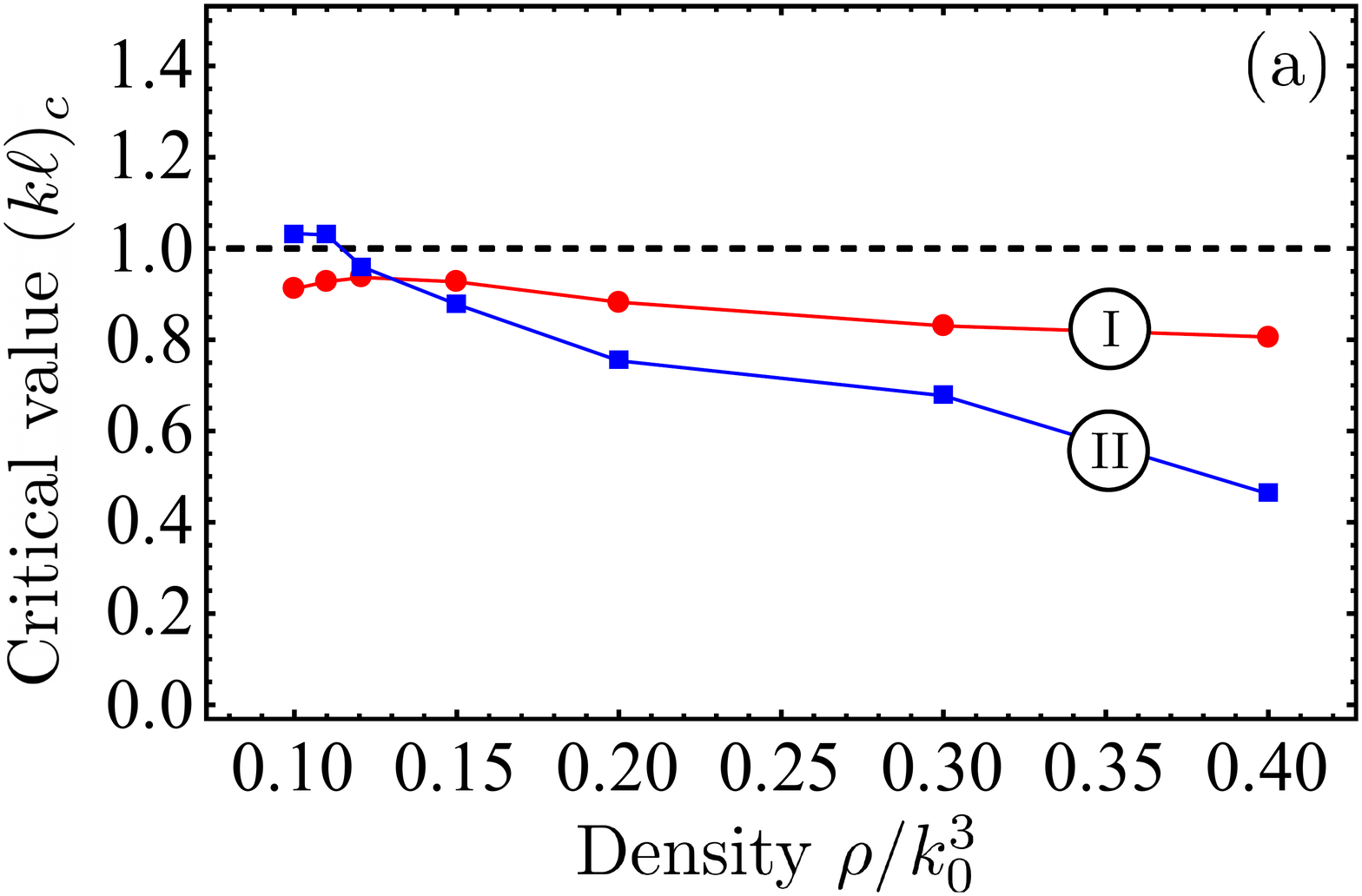}
%% \;\;\;\;
\includegraphics[width=0.99\columnwidth]{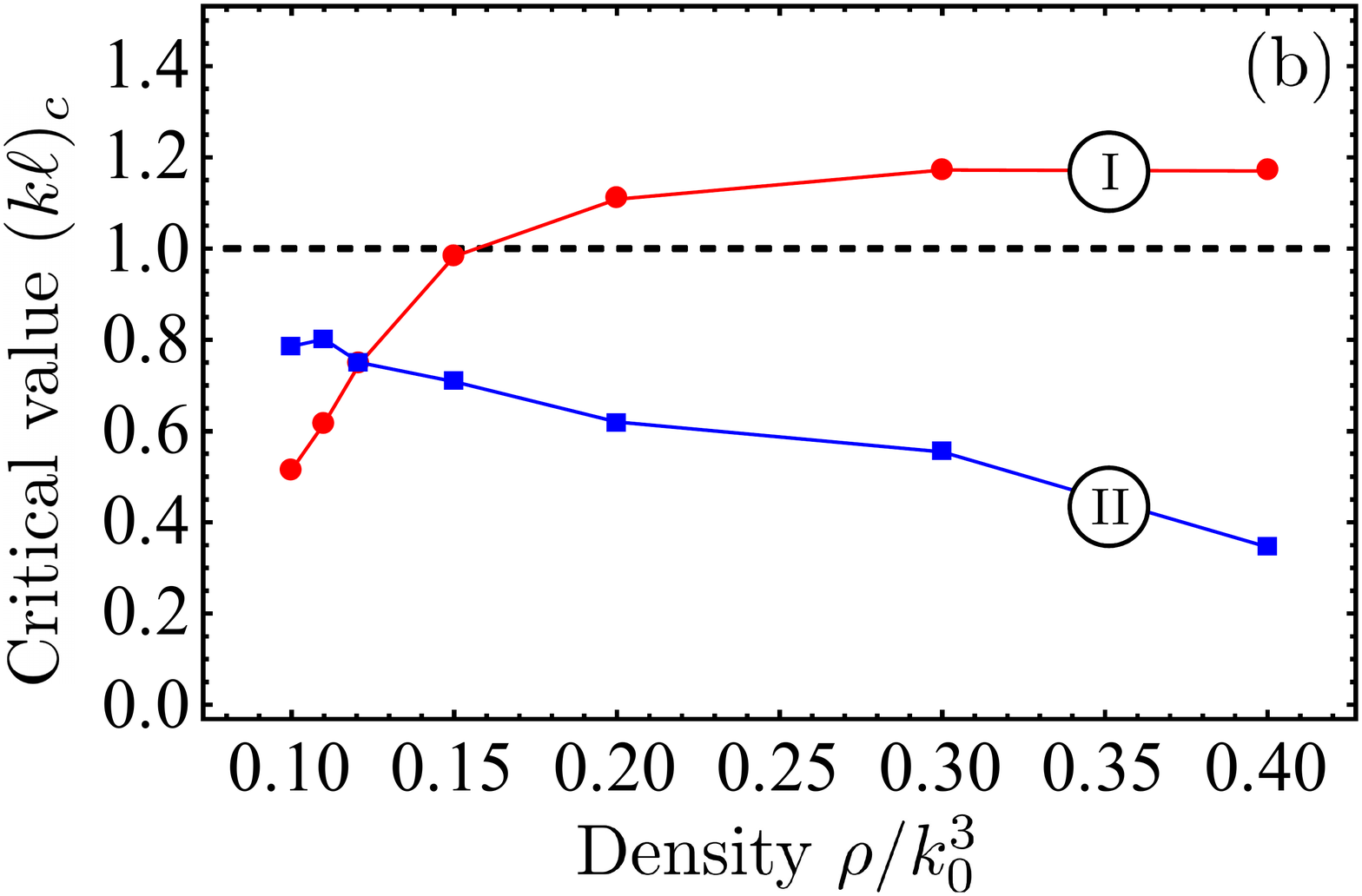}
%%\vspace*{-1.3cm}
\caption{\label{fig_klc} Values of the Ioffe-Regel parameter $k \ell$ at the mobility edges $\omega =  \omega_c^{\mathrm{I, II}}$ as functions of scatterer number density $\rho/k_0^3$ (low-frequency mobility edge I---red circles, high-frequency mobility edge II---blue squares). (a) $\ell$ and $k$ were estimated by the integral formulas (\ref{integralell}) and (\ref{integralk}) with $k_0 z_{\mathrm{max}} = 4$. (b) $\ell$ and $k$ were estimated from linear fits of Eqs.\ (\ref{az}) and (\ref{phiz}) to the numerical data in the interval $k_0 z \in (1, 4)$. The horizontal dashed lines show $(k \ell)_c = 1$ expected at a mobility edge.}
\end{figure}

Another way to determine $\ell$ and $k$ is to enforce the behavior dictated by Eqs.\ (\ref{az}) and (\ref{phiz}) and determine $\ell$ and $k$ as best fit parameters from linear fits of  Eqs.\ (\ref{az}) and (\ref{phiz}) to $\ln A(z)$ and $\varphi(z)$, respectively. The fits are to be applied in a certain range of depths $z \in (z_{\mathrm{min}}, z_{\mathrm{max}})$ that we can choose to minimize the impact of any undesirable artifacts such as, e.g., the proximity of a boundary at $z=0$ or the bad quality of numerical data at large $z$. The definitions (\ref{integralell}) and (\ref{integralk}) are mainly sensitive to the behavior of the average wave field $\langle \psi(z) \rangle$ at short distances because $A(z)$ in Eq.\ (\ref{integralell}) decays rapidly with $z$ whereas the integration in Eq.\ (\ref{integralk}) is explicitly restricted to $z < z_{\mathrm{max}}$. In contrast, the fits of $\ln A(z)$ and $\varphi(z)$ by linear functions may account for the behavior of $\langle \psi(z) \rangle$ in a wider range of depths depending on the choice of the fit interval $(z_{\mathrm{min}}, z_{\mathrm{max}})$.

\section{Calculation of the Ioffe-Regel parameter and discussion of results}
\label{ir}

We now apply the two methods of determining $\ell$ and $k$ to the numerical data of Fig.\ \ref{fig_ampl_phase} and calculate the resulting Ioffe-Regel parameter $k\ell$ at the mobility edges I and II, as a function of scatterer density $\rho$. The results are shown in Fig.\ \ref{fig_klc}. For the considered range of densities $\rho/k_0^3 = 0.1$--0.4, the critical values of $k\ell$ vary between  {0.3} and  {1.2}. This demonstrates that even if the criterion $(k\ell)_c = 1$ turns out to be qualitatively valid to determine the positions of mobility edges (because  {0.3} and  {1.2} are still of order 1), it is far from having a strict quantitative validity. For both definitions of $k$ and $\ell$, the best agreement between the critical values of $k\ell$ at the two mobility edges is achieved at low densities $\rho/k_0^3 = 0.1$--0.2, whereas high-density values of $(k\ell)_c$ start to differ significantly, by almost a factor of  {4} in Fig.\ \ref{fig_klc}(b).

It is instructive to compare our results for scalar waves with those obtained previously for light with account for its vector character. A calculation of coherent light propagation analogous to the one presented in Sec.\ \ref{coherent} yields a purely exponential decay of the wave amplitude $A(z)$ and a linear growth of its phase $\varphi(z)$ in the range of densities $\rho/k_0^3 = 0.1$--0.5 and for all frequencies $\omega$ near the resonance frequency $\omega_0$ \cite{fofanov11,note2}. Thus, both the mean free path $\ell$ and the effective wave number $k$ are well defined, leading to the values of the Ioffe-Regel parameter $k\ell$ as small as 0.5 \cite{fofanov11}. However, despite the fact that this value is of the same order as or even smaller than the critical values $(k\ell)_c$ reported in Fig.\ \ref{fig_klc}, the eigenvectors of the corresponding Green's matrix remain extended and no Anderson localization takes place \cite{skip14}. This appears rather counter-intuitive in the light of the available analytical results \cite{bart94,cherroret16}. Indeed, the calculation of the self-energy $\Sigma$ shows that $\Sigma$ becomes momentum-dependent already in the second order in density $\rho/k_0^3$ for both scalar waves \cite{bart94} and vector light \cite{cherroret16}. However, the results of Ref.\ \onlinecite{fofanov11} suggest that for light, the momentum-dependent part of $\Sigma$ remains always small and can be neglected. The reason for this is unclear at the moment. The remaining momentum-independent part of $\Sigma$ yields well-defined mean free path $\ell$ and effective wave number $k$ of the optical wave in the scattering medium. In contrast, for scalar waves the momentum-dependent part of $\Sigma$ becomes significant at high densities of scatterers, which creates difficulties for defining $\ell$ and $k$ properly (see Sec.\ \ref{coherent} and, in particular, Fig.\ \ref{fig_ampl_phase}). Altogether, it turns out that the model of scalar wave scattering by an ensemble of resonant point scatterers is much richer than its vector (optical) counterpart because it contains both different regimes of coherent wave propagation (with and without well-defined $\ell$ and $k$) and different regimes of transport by multiple scattering (diffusion and Anderson localization). In contrast, the vector optical model always yields well-defined $\ell$ and $k$ \cite{fofanov11} and exhibits no Anderson transition \cite{skip14}.

\section{Conclusions}
\label{concl}

In this work, we established a phase diagram for a scalar wave in an ensemble of resonant point scatterers. Localized modes appear at sufficiently high number densities of scatterers  $\rho > \rho_c$ in a wedge-shaped region on the density-frequency plane. The critical density $\rho_c$ is estimated to be $\rho_c/k_0^3 \simeq 0.08$, which corresponds to $k_0 \ell_0  = 1$ with $\ell_0$ the on-resonance mean free path in the independent-scattering approximation. The propagation of the coherent component of an incident wave with a frequency at the mobility edge is strongly affected by the spatial dispersion of the effective medium, resulting in a nonexponential decay of the amplitude $A(z)$ of the average wave field and a nonlinear growth of its phase $\varphi(z)$. This makes impossible the definition of the mean free path $\ell$ and of the effective wave number $k$ in a usual way, as the characteristic length of the exponential decay of the wave intensity $A(z)^2$ and the rate of the linear growth of the phase $\varphi(z)$, respectively. Despite this difficulty, we defined $\ell$ and $k$ as {\em effective} decay  {length of $A(z)^2$ and growth rate of $\varphi(z)$}, respectively. This allowed us to calculate the value $(k\ell)_c$ of the Ioffe-Regel parameter $k\ell$ at the two mobility edges as a function of scatterer density and show that it takes values from  {0.3} to  {1.2}. Hence, the usual form of the Ioffe-Regel criterion $k\ell < (k\ell)_c = \mathrm{const} \sim 1$ is a qualitatively correct but quantitatively inexact condition of Anderson localization in 3D even for such a simple model of disordered medium as a random ensemble of  {resonant} point scatterers.
 {Despite its quantitative inaccuracy, the Ioffe-Regel criterion turns out to be much more relevant as a criterion of Anderson localization for resonant scatterers than for disordered media with nonresonant scattering, such as, for example, the mechanical systems studied in Refs.\ \onlinecite{sheng94,beltukov13,beltukov17}, where the Ioffe-Regel frequency $\omega_{\mathrm{IR}}$ at which the condition $k\ell = 1$ is obeyed can be very different from the localization transition frequency $\omega_c$. At the same time, the localization transitions in systems with resonant and nonresonant scattering belong to the same universality class \cite{skip16,beltukov17}.}

 {The model considered in this work accounts for the resonant nature of scattering in disordered systems used in experiments on Anderson localization of sound \cite{hu08,cobus16} or light \cite{wiersma97,vanderbeek12,storzer06,sperling13}. Being near a scattering resonance is important to obtain a localization phase diagram with two mobility edges as the one shown in Fig.\ \ref{fig_me}. In this respect, our model reproduces the behavior found in the acoustic experiment of Ref.\ \onlinecite{cobus16} where a band of localized states (a mobility gap) was found between two mobility edges that are close to but do not coincide with a single-scatterer resonance frequency. On the other hand, our model does not account for the finite scatterer size and correlations in scatterer positions. The impact of the latter correlations on the phenomenon of Anderson localization is under an active study in systems of dimensionality larger than one \cite{rojas04,froufe17}. Correlation in scatterer positions can be readily taken into account in our model of point-like scatterers, which was already used to study systems ranging from weakly, short-range correlated (e.g., a minimum distance between scatterers is imposed \cite{caze10}) to aperiodic deterministic (e.g., complex prime arrays \cite{wang18}). In contrast, accounting for a finite scatterer size $\sim \lambda$ turns out to be much more complex and requires implementation of advanced numerical methods \cite{mish07,conti08}. In both cases of correlated scatterer positions or finite scatterer size, the parameter space of the problem in extended to at least one additional dimension (e.g., the degree of correlation or the scatterer size), which makes it difficult to explore entirely. Such an exploration may be a subject of future work.}

Thus, there is still no simple and reliable {\em quantitative} criterion of Anderson localization that would be suitable to guide experiments. It is nevertheless worthwhile to note that the model discussed in this work could be a good testbed for eventual new criteria of localization because its phase diagram is now known (see Fig.\ \ref{fig_me}) and can be recalculated with arbitrary precision if needed. Correlations in scatterer positions or vector nature of considered waves (e.g., light \cite{skip14} or elastic waves \cite{bel18}) can be readily incorporated into the model.

\acknowledgements
This work was funded by the Agence Nationale de la Recherche (project ANR-14-CE26-0032 LOVE). The analysis of the coherent wave field was performed with a financial support of the Russian Science Foundation (grant 17-12-01085). A part of the computations presented in this paper were performed using the Froggy platform of the CIMENT infrastructure ({\tt https://ciment.ujf-grenoble.fr}), which is supported by the Rhone-Alpes region (grant CPER07\verb!_!13 CIRA) and the Equip@Meso project (reference ANR-10-EQPX-29-01) of the programme Investissements d'Avenir supervised by the Agence Nationale de la Recherche.


\begin{thebibliography}{99}
\bibitem{note1}
In optical experiments, one most often measures the transport mean free path $\ell^*$ and not $\ell$, but we expect $\ell$ and $\ell^*$ to be of the same order for materials of Refs.\ \cite{wiersma97,vanderbeek12,storzer06,sperling13}.
\bibitem{note2}
In Ref.\ \onlinecite{fofanov11}, $\ell$ and $k$ were determined from the analysis of the average atomic polarization induced by a circularly polarized plane wave incident on an ensemble of two-level atoms ($J =0 \to J = 1$ transition).
\end{thebibliography}
\end{document}